\documentclass[useAMS,usenatbib]{article}
\usepackage{times}
\usepackage{bm}
\usepackage{natbib}
\usepackage[figuresright]{rotating}
\usepackage{graphicx}
\usepackage{amssymb}
\usepackage{amsmath}
\newcommand{\BB}{{\cal B}}

\newcommand{\Ee}{{{\rm E}}}
\newcommand{\ob}{{{\rm obs}}}
\newcommand{\ex}{{{\rm ex}}}

\newcommand{\dd}{{\rm d}}

\newcommand{\PP}{{{\cal P}}}

\usepackage{colortbl}

\title{Causality without potential outcomes and the dynamic approach}

\author{Daniel Commenges
INSERM, U 1219, Bordeaux,  F33076, France}

\begin{document}





\label{firstpage}



\maketitle

\begin{abstract}
Several approaches to causal inference from observational studies have been proposed. Since the the proposal of \citet{rubin1974estimating} many works have developed a counterfactual approach to causality, statistically formalized by potential outcomes. \citet{Pearl2000} has put forward a theory of structural causal models which gives an important role to graphical models and do not necessarily use potential outcomes. On the other hand several authors have developed a dynamical approach in line with \citet{Granger1969}. We analyze prospective and retrospective causal questions and their different modalities. Following \citet{Dawid2000} we develop criticisms about the potential outcome approach and we show that causal effects can be estimated without potential outcomes: in particular direct computation of the marginal effect can be done by a change of probability measure. Finally, we highlight the need to adopt a dynamic approach to causality through two examples, ``truncation by death'' and the ``obesity paradox''.
\end{abstract}\vspace{5mm}

{\bf keywords:} causality; obesity paradox; inverse probability weighting; potential outcomes; truncation by death; stochastic system

\section{Introduction}\label{sec:Introduction-DynCaus}
Since it was proposed by \citet{fisher1935design}, the method of randomized trials has been a very important tool for establishing causality. The need for a formalism with greater generality and applicable to observational studies, however, has appeared.
 The use of potential outcomes has first been proposed by Jerzy Neyman in 1923  \citep{Splawa-Neyman1990}. The idea was revived and formulated in modern notation by \citet{rubin1974estimating}. Since then, causal inference based on potential outcomes has been developed in a series of papers and has become the dominant school of causal inference in Biostatistics and is also influent in other fields such as econometrics \citep{Heckman2005a}. This theory was reviewed by \citet{holland1986statistics} and \citet{rubin2005causal} among others. Robins and coworkers  used the potential outcomes approach to more complex problems, in particular to the problem of dynamic treatment regimes \citep{Robins2000,robins2009estimation}.

 \citet{Pearl2000,Pearl2009} has put forward a theory of structural causal models which gives an important role to graphical models and do not necessarily use potential outcomes.

 On the other hand several authors have attempted to approach causality through dynamical models. This has been first proposed by  \citet{Granger1969} working with time series, and has been given a more powerful formalism by the works of \citet{Schweder1970} and \citet{Aalen1987}, further developed by \citet{arjas2004causal} and \citet{Didelez2008}. This approach is particularly well adapted for studying feedback and mediation \citep{Aalen2012a,aalen2018feedback}. Of course the use of a dynamical model is not sufficient for causal interpretation but it is possible to formalize the assumption needed, particularly through the concept of ``system'' elaborated in \citet{Commenges2009} and in Chapter 9 of \citet{commenges2015dynamical}. This stochastic system approach to causality has similarity with Pearl's structural causal models with the important difference that it is based on stochastic processes rather than simple random variables.

 The potential outcomes (or counterfactual) approach has been criticized by \citet{Dawid2000} and the title of this paper echoes Dawid's paper. Indeed, it seems to us that the approach using potential outcomes is not in line with the classical statistical and probability theory and, partly because of this fact, raises a number of problems, both theoretical and practical. It is not in line either with the concept of causality implicit in physics. One aim of this paper is to review the different criticisms that can be addressed to this approach. Some of them have already been raised by \citet{Dawid2000} and \citet{Aalen2012a} and some of them are new. The other aim is to put forward the need of a dynamic approach to causality.

The organisation of the paper is as follows. In Section 2 we examine different types of causal questions, distinguishing between prospective and retrospective questions. In Section 3 we highlight several problematic issues about potential outcomes. In Section 4 we show that conditional and marginal effects can be defined and computed without resorting to potential outcomes. In Section 5 we put forward the need  of a dynamic approach to causality and we illustrate its potentiality by two applications: treating the problem of truncation by death and analysing the obesity paradox. Section 6 concludes.

\section{The causal questions}\label{sec:causalquestions}

\subsection{Generalities}
The ``potential outcomes'' concept is closely linked with that of ``counterfactuals''; we must first define (and distinguish) the two concepts.
Potential outcomes are attributed to each subject according to the value of a treatment variable. For instance in the case of a binary treatment, each subject has two potential outcomes $Y_i(0)$ and $Y_i(1)$ for treatments $0$ and $1$ respectively. Once the study is done, one of these variables is ``counterfactual'' since subject $i$ received either treatment $0$ or $1$; if subject $i$ received treatment $1$, it goes ``against the fact'' to say that he received treatment $0$; this means of course that $Y_i(1)$ is observed but not $Y_i(0)$. While in this approach, the treatment effect for subject $i$ is defined as $Y_i(1)-Y_i(0)$, it is clear that it cannot be computed because only one the terms can be observed. This was called by \citet{holland1986statistics} ``the fundamental problem of causality''. Thus the use of the potential outcomes implies considering counterfactuals, but not all potential outcomes are counterfactual. Also we should distinguish counterfactual events and counterfactual questions.

\subsection{Prospective (Q$_p$) and retrospective (Q$_r$) questions}
In philosophy of science, an important distinction  has been made between ``general causation'' and ``singular, token'' or ``actual causation''. \citet{Hitchcock} writes: ``General causal claims, such ``smoking causes lung cancer'' typically do not refer to particular individuals, places, or times, but only to event-types or properties. Singular causal claims, such as ``Jill’s heavy smoking during the 2000s caused her to develop lung cancer'', typically do make reference to particular individuals, places, and times.''
This distinction, however, is not precise enough because it mixes two different oppositions: prospective/retrospective and general/singular.
In fact, the so-called ``general causation'' questions are prospective, while token causation questions are retrospective.
\citet{Dawid2000} called the two types of questions ``the effects of cause'' and the ``the cause of the effect'' and also noted that some effect of cause (prospective) questions were in fact ``singular''; \citet{pearl2015causes} also used this terminology. Since cause precedes effect, this opposition is congruent with the prospective/retrospective opposition.
The opposition between ``general'' and ``singular'' is more a question of level of analysis: for instance individual or population level (see Section \ref{sec:modalities}). A prospective question can be at the individual level, when the particular characteristics of an individual are included in the model; personalized medicine, for instance, attempts to make individual predictions and treatments \citep{chakraborty2014dynamic}.

A typical prospective question (effects of cause), denoted Q$_p$, at the individual level is: ``What will be the difference in the outcome between the two actions: giving treatment $0$ or giving treatment $1$ to subject $i$ ?'' This can be answered naturally, without resorting to potential outcomes, by comparing the distributions of the outcome $Y$ under two different probability laws $\PP_0$ and $\PP_1$ obtained for treatment $0$ and $1$ respectively, in a situation where we can manipulate the treatment (as in an experiment);  for instance, we can contrast them by considering $\Ee_1(Y)-\Ee_0(Y)$. We have of course to estimate $\PP_0$ and $\PP_1$ from observations. Q$_p$ is prospective, so does not imply counterfactuals, since the fact has not yet been observed.

A typical retrospective question (cause of effect), denoted Q$_r$ , is: ``Subject $i$ has experienced event $Y=1$; what is the cause of this event''. It may happen that the question is focused on a particular factor, for instance the treatment he took; assume that he took treatment $0$. One wonders whether taking treatment $0$ took a role in the subject experiencing $Y=1$. If we know the probabilities of event $Y=1$ after taking treatment $0$ and $1$, respectively $\PP_0(Y=1)$ and $\PP_1(Y=1)$, we can contrast them, using either a difference or a ratio. As for answering to Q$_p$, we have in practice to estimate $\PP_0$ and $\PP_1$ from observations. If $\PP_0(Y=1)>\PP_1(Y=1)$, we may think that taking treatment $0$ rather than $1$ is one of the causes of $Y=1$ for subject $i$; of course the magnitude of the contrast between $\PP_0(Y=1)$ and $\PP_1(Y=1)$ plays a role in the causal attribution. Here the choice of the contrast is important (example of lung cancer and smoking where a ratio indicates a large effect while a difference gives a modest effect). Q$_r$ is retrospective. It has a more counterfactual flavour than Q$_p$ since the facts are already there and we can ask ``what if'' subject had taken treatment $1$ rather than $0$ ? However, we see that we can answer this counterfactual question by simply computing $\PP_1(Y=1)$ and $\PP_0(Y=1)$, that is without resorting to potential outcomes.

\citet{arjas1999probabilistic} has solved some intriguing riddles in retrospective (token) causation by resorting to latent variables.

\subsection{Modalities of the Q$_p$ and Q$_r$ questions: individual vs population levels, multiplicity}\label{sec:modalities}
Both types of questions have several modalities. First they can pertain to the individual or to the population level, and these levels are in interaction. We have formulated the questions above at the individual level, but we can formulate them also at the population level. Assume for simplicity that we can manipulate the prevalence of smoking as if it was a treatment; the outcome of interest is the incidence of lung cancer in a given population. The answer to the population-level Q$_p$ question, the effect of smoking prevalence on incidence of lung cancer, can be deduced from the effect of smoking at the individual level (although it involves here a rather complicated dynamic model).
In practice it is not possible to manipulate smoking prevalence but it may be possible to modify it via an anti-tobacco campaign, which is really a ``treatment'' at the population level; the effect of an anti-tobacco campaign on lung cancer incidence is mediated by the smoking prevalence. This example raises the issue of the difference between the causal effect of a ``state'' and the causal effect of an intervention, well described by \citet{pearl2018does,pearl2019interpretation} at the individual level.

In fact, the research on tobacco started with a Q$_r$ question for lung cancer. The question, asked in the 1950's \citep{breslow1954occupations}, was ``what is the cause of the increase in lung cancer incidence ?''. Of course, this question was asked because descriptive studies had observed the fact of increasing incidence. Studies for finding risk factors at individual level  were then launched; they showed that smoking was an important risk factor of lung cancer; moreover, it was also observed that smoking prevalence had increased; it was deduced that the increase in smoking prevalence in the beginning of the 20$^{th}$ century was responsible of the increase of lung cancer incidence in the second part of the same century. This illustrates both the relation between individual and population levels and Q$_p$ and Q$_r$ questions.

The other type of modalities arises from the possible multiplicity of effects and of causes for Q$_p$ and Q$_r$, respectively. An event may have several effects: an example of the multiple Q$_p$ is the study of the effects of  hormonal therapy on several diseases such as cardiovascular diseases or cancer \citep{writing2002risks}. In clinical trials  a primary outcome is specified, but there are often secondary outcomes, and  also, side effects are examined.
On the other hand, an event may have several causes. In the case of lung cancer (a Q$_r$ question), one major cause was found; this does not exclude the existence of other factors and it is not always the case that one factor has a much larger impact than any other factor. An example of multiple Q$_r$ is given by a study on Alzheimer's disease. Epidemiologists record the occurrence of the disease and also many factors which may possibly be risk factors; low educational level and some genetic factors (like APO E4 allele) have been found. Such studies allow us to learn the law of the system, that is the conditional probability $\PP(Y|Z)$ where $Z$ are the possible causes; in simple analyses, $Y$ is a random variable representing the disease of interest, $Z$ is a vector of ``explanatory'' variables; in more elaborated analyses, the disease and part of the potential risk factors may be represented by stochastic processes as will be discussed in Section \ref{sec:SSAC}. Of course in any case, for going from simple association to causal interpretation, much care must be exercised, particularly to avoid conventional or dynamic confounding.

\section{Problems with potential outcomes}\label{sec:potentialoutcomes}
As was said in the Introduction Section, many recent developments in the formalism of causality have used the potential outcome approach. We develop here a criticism of this approach.
\subsection{A list of issues with the potential outcome approach}
\begin{enumerate}
\item It is at odds with the Kolmogorov formalism and it may raise theoretical problems with continuous variables and continuous time
\item The status of random variables for the potential outcomes is not clear
\item The causal effect is defined in the sample while we want to extrapolate
\item There is most often no modeling of measurement error
\item The whole approach is influenced by randomized trials for one treatment while in epidemiology causality is multiple
\item Most of the work focus on marginal effects while conditional effects are very important
\item Time is not fully taken into account
 \end{enumerate}

As a consequence of these shortcomings, the potential outcome approach has difficulties for treating important causal problems such as mediation, truncation by death (Section \ref{sec:truncationbyD}), dynamic selection (Section \ref{sec:DynSelec}) and feedback (Section \ref{sec:feedback}).

\subsection{Points 1, 2 and 3: the probabilistic issues}
The theoretical foundation of probability dates back from Kolmogorov (1933). In this formalism, a random element is defined as an application from a space $\Omega$ equipped with a sigma-algebra (or sigma-field), say $\Sigma$, to another space, also equipped with a sigma-algebra. The sigma-algebra specifies which events an be given a probability. Random elements are essentially random variables or stochastic processes. First consider the case of a real random variable, $Y$. As presented by \citet{williams1991probability}, $Y$ is an application from the ``universe'' $(\Omega,\Sigma)$ on  to the real line $(\Re,\BB)$, where $(\Omega,\Sigma)$ and $(\Re,\BB)$; for each element $\omega\in \Omega$  the random variable takes the value $Y(\omega)$. Thus, in a sense, a random variable has already potential outcomes! It is important to realize that there is no need to specify a probability measure to specify the random elements: the random variable only specify which elements of $\Re$ are associated to elements of $\Omega$. Then, on these measurable spaces we can put probability measures. A probability measure on $(\Omega,\Sigma)$ induces a probability measure on $(\Re,\BB)$ through the variable $Y$, which is called the distribution of $Y$. One can assume that there is a true probability measure according to which ``Nature'' chooses $\omega$, which induces the true distribution of $Y$. Since what we can call the ``fundamental problem of {\em statistics}'' is that we do not know the true probability, we choose a family of probability measures (or of distributions) in which we hope that the true probability measure lies; this is what is called a ``statistical model'', and if the hope is fulfilled, it is well-specified \citep{commenges2009statistical}. Then, the statistical theory of inference allows to choose a distribution, based on observations, that is close to the true distribution. All this is now conventional, but note that Neyman proposed the potential outcome notation in 1923, before Kolmogorov proposed his formalism. When using this formalism in presence of an explanatory variable $A$, conventional statistics uses conditional probabilities, and for instance for a binary variable $A$, $Y$ has the distribution $\PP_0$ if $A=0$ and $\PP_1$ if $A=1$. Of course this does not make a ``causal model'', but this is the corner stone of the basic formalism that can be used for this aim. The fact that ``correlation is not causation'' has been acknowledged for a long time in statistics and epidemiology, and the concept of ``confounding factor'' has long been a key concept in epidemiology.

Let us now examine the potential outcome formalism. In the formalism proposed by Neyman and rejuvenated by \citet{rubin1974estimating}, the potential outcomes are fixed. Each subject has a potential outcome $Y_a$ which is observed if he takes the treatment $A=a$. The random elements are the treatment attribution or the label of the units. This is clearly described in \citet{rubin2005causal}. So, the potential outcomes does not seem to be random variables. The effects are defined  by $\Ee(Y_a)$, where the expectation represents a mean in the population.

\citet{neyman1935statistical}  proposed design-based analysis using potential outcomes. Note however that this can be done without potential outcomes. A particular type of design-based analysis is analysis by randomisation tests studied by \citet{pitman1937significance}. In randomization tests the distribution of the test statistic used is conditional on the outcome of interest and is generated by the randomization itself. Design-based analysis such as in \citet{lemeshow1998illustration} also allows inference based on design in observational studies.
Limitations of design-based analysis are: (i) that we have to extrapolate to the general population and (ii) more importantly, we have to specify a statistic. The theory does not provide natural statistics but only yields the distribution of pre-specified statistics; by contrast, the likelihood theory provides estimators and test statistics with good properties.

\subsection{Point 4: difficulty with modeling measurement error}
In spite of  a thorough analysis of ``measurement bias'' in Chapter 9 of \citet{Hernan2018causal}, and recognising that assuming no measurement error is unrealistic, the methods of causal inference based on potential outcome most often assume that there is no measurement error: see \citet{Hernan2018causal} Chapter 13.5. Note, however, that Neyman (1935) introduced ``technical errors''.

\subsection{Point 5: randomized trials for one treatment vs multiple causality}

\citet{Hernan2018causal} in Chapter 3 and Section 13.5 state that
the analysis of observational data should emulate that of a
hypothetical randomized experiment as closely as possible.
In most works using potential outcome the question is to assess the causal effect of a treatment of a particular factor (like smoking). Moreover the values of the treatment to be compared correspond to well defined interventions (consistency). As important as it is, this problem does not cover all fields of application of causal inference. In epidemiology there are often several factors which influence the occurrence of a disease in a complex way involving interactions. Moreover, it is possible to speak of causal effect in cases where no intervention seem possible (effect of the moon on tides, effect of genetic variants), or difficult (BMI). For BMI there is a cascade of effect: ``recommendation of diet'' --$>$ ``effective diet'' --$>$ BMI --$>$ mortality (see \citet{pearl2018does} for the effect of BMI contrasted with the effect of an intervention). Also the factors of interest are often continuous and the effect shape may be complex (for instance U or J shape for the effect of BMI on death rate).

\subsection{Point 6: focus on marginal effects vs  conditional effects}
The focus of most of the literature using potential outcomes has been on marginal effects. This is probably not an absolute limitation of the approach. For instance, the marginal structural models have been extended to fixed covariates \citep{Cole2005}. However the approach has not been used for analyzing complex effects of several factors as may be required for developing personalized medicine. Moreover, the potential outcomes approach does not seem to be able to incorporate random effects.

\subsection{Point 7: time not fully taken into account}
In a large body of the literature using potential outcome, time is only implicitly taken into account. In simple contexts such as those of clinical trials where a treatment $A$ is given and an outcome $Y$ is defined, it is implicit that $A$ is given before the outcome $Y$ is observed and this may be sufficient in this case. However, in more complex models there are several variables which are implicitly ordered and there may be direct and indirect effect. \citet{Aalen2012a} have shown how a model including time is superior to a model which does not. One of the example is taken from \citet{cole2002fallibility}. Also \citet{aalen2016can} showed that discretization of time could lead to erroneous causal interpretations. The possibility of a dynamic approach will be discussed in Section  \ref{sec:SSAC}.

Also, as analyzed by \citet{aalen2019time}, in case of mediation, the natural direct effects and the natural indirect effects constructed with potential outcomes are ill-defined in a survival context because the manipulation of the mediator at time $t$  may make no sense if the subject is dead or censored before this time.

\section{Causality without potential outcome in the random variables context}\label{section:withoutPO}
\subsection{The do operator and its probabilistic interpretation}
\citet{Pearl2000} has developed structural causal models  (SCM) in which the causal effect of a factor is defined by the "do operator". The do-operator is defined graphically by cutting all the arrows pointing toward this factor. The aim is to define the effect of the factor in a situation where it is not influenced by any other factor, as in a randomized experiment.

\begin{figure}[h!]

\centering
\includegraphics[scale=0.45]{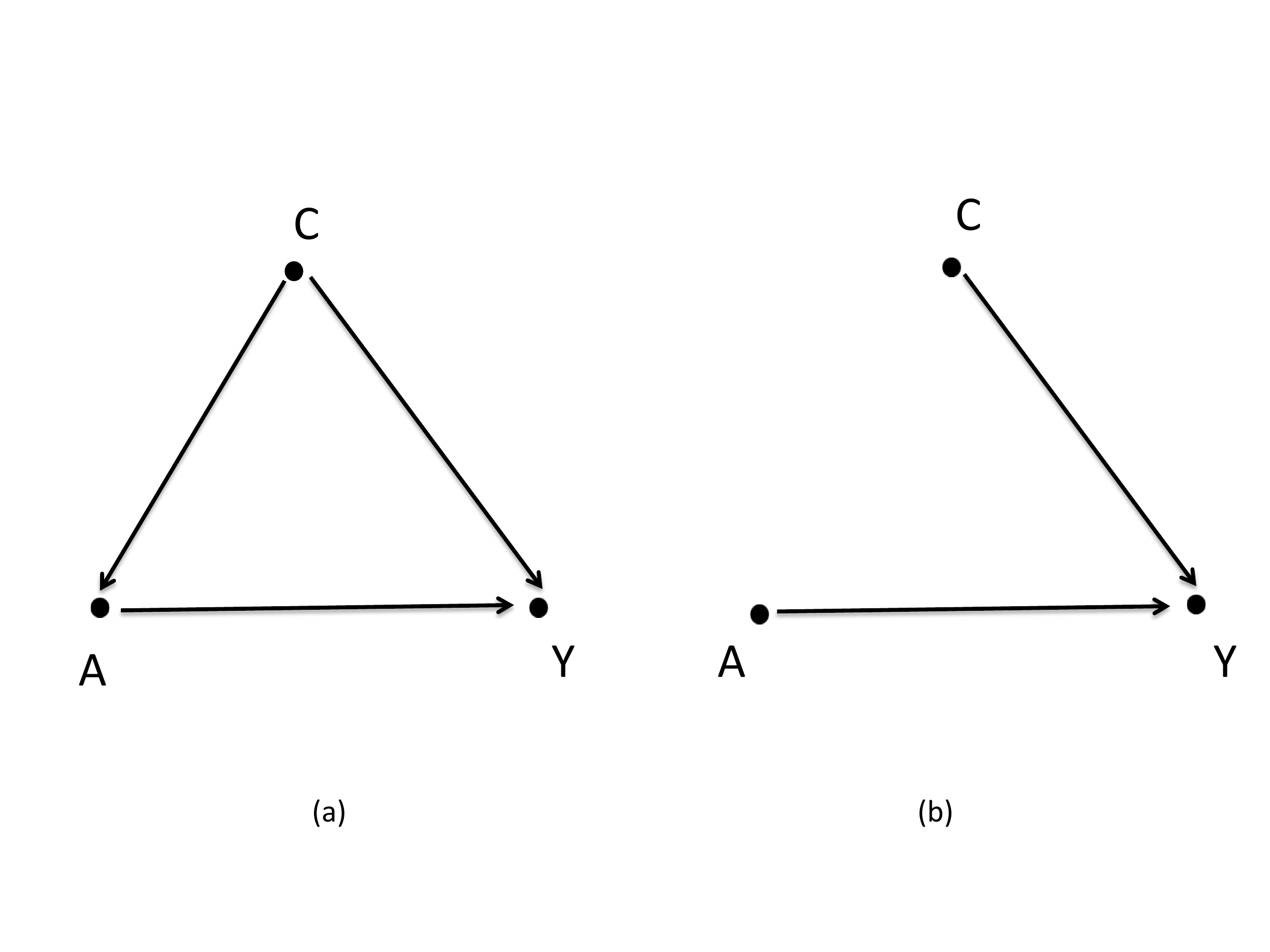}
\caption{Influence graphs for $A$, $Y$ and $C$: a) observational study; b) experiment.\label{graph-causal-obs}}
\end{figure}

For illustration, we consider the following simple setting depicted in Figure \ref{graph-causal-obs} (a): there is a confounding factor $C$ which influences both $Y$ and $A$. We assume that this is the only confounding factor (possibly multidimensional); that is, the situation is that of no unmeasured confounders if we include $C$ in the system. The graph of Figure \ref{graph-causal-obs} (b) is that obtained by the do operator  do A=a. The effect of $A$ is defined by $\Ee (Y|do A=a)$  This setting is described in Figure \ref{graph-causal-obs} b) ; it is similar to the ``do operator'' of \citet{Pearl2000}. We do not need to invoke counterfactuals to describe it; this is rather a conceptual situation, which may also be realized in the future; we do not need either to use potential outcomes to analyse this setting. In this experimental setting, the law of the three variables $Y,A,C$ is given by a probability measure $\PP^{\ex}$; in this law, the conditional distribution $f_{Y|A,C}$ and the marginal distribution of $C$, $f_C$, remain the same as in the observational setting, but $C$ does not influence $A$, that is $f^{\ex}_{A|C}=f^{\ex}_{A}\ne f_{A|C}$.
 This approach has been formalized by \citet{arjas2012causal}.

\subsection{Conditional and marginal effects}

It is worth to note that both the researchers engaged in the potential outcome approach and Pearl have emphasized the importance of the marginal effect. This is $\Ee (Y|do A=a)$ in in Pearl' SCM and $\Ee(Y_a)$ in the potential outcome approach ; contrasts for different values of $A$ and other functional of the distribution can be considered. The effect thus defined is marginal. The conditional effects, however, appear to be more and more important with the development of personalized medicine.

\citet{Pearl2000} has studied the identifiability of the effect. The back-door and the front-door criteria (not useful in our simple example) allow defining sets of variables which are sufficient for computing the effect $\Ee (Y|do A=a)$. Our point here is to note that the SCMs do not use potential outcome and that there is a probabilistic interpretation of the do-operator. In fact a graph is a representation of properties of the distributions of the variables in the system (here, conditional independence). So, for instance in our example of Figure \ref{graph-causal-obs}, we are just contemplating the influence graphs of the system $(Y,A,C)$ under a probability measure which is that of an observational study for graph (a) and that of an experiment in graph (b). In our example we would note the effect of ``doing $A=a$'' $\Ee_a (Y)$, meaning that we compute the expectation under the probability measure where we have fixed $A$ to the value $a$; this can also be noted $\Ee_{ex}(Y|A=a)$, where $\Ee_{ex}$ means expectation under an experimental measure (where $A$ is not influenced).

\subsection{Computation of marginal effects through conditional effects}\label{sec:throughcond}
The computation is of course the same as with the do operator: $\sum_c \Ee_a (Y|C=c)P(C=c)$; both quantities $\Ee_a (Y|C=c)$ and $P(C=c)$ can be estimated from an observational study.

A formula valid when $Y$ and $C$ are continuous is:
\begin{equation}\label{margeffect} \Ee_{\ex}[Y|A]=\int y f_{Y|A,C}(y|A,c) f_C(c)~~ \dd y ~~\dd c.\end{equation}
  These formulas are related to the ``standardized mean'' in epidemiology and  called the g-formula by \citet{robins2009estimation}. In an observational study, $f_{Y|A,C}$ has to be estimated.

There is nor potential outcome, nor counterfactual in this point of view. If $A$ is manipulable, the probability measure corresponding to "do A=a" can be realized; if it is not, it can still be considered from a mathematical point of view \citep{pearl2019interpretation}.


\subsection{Computation of the marginal effects by inverse probability weighting}\label{sec:margeffect}

Here we show in our simple setting, using random variables, how causal analysis using IPTW can be derived without potential outcomes.
With the idea to mimic a randomized trial, it has been proposed to directly estimate the marginal effect, without estimating $f_{Y|A,C}$. If $A$ is discrete, the marginal expectation of $Y$ for $A=a$, $\Ee[Y_a]$, is simply estimated by a weighted mean of $Y$ using unstabilized or stabilized weights, $W(A,C)=\frac{1}{f_{A|C}(A,C)}$ and $W_S(A,C)=\frac{f_A(A)}{f_{A|C}(A|C)}$, respectively \citep{hernan2006estimating}; it is claimed (under positivity and exchangeability assumptions) that $\Ee(Y_a)=E[W(A,C)I(A=a)Y]$. However if $A$ has a continuous distribution this expression is problematic. Of course, in observational studies, $f_A$ and $f_{A|C}$ have to be estimated.

If $A$ is continuous (or can take a large number of values), a model is assumed for the marginal effect of $A$ and an estimation equation is proposed. In the potential outcome formalism this is noted $\Ee_{\beta}[Y_a]=g(a;\beta)$. Assuming the marginal model well-specified, a weighted estimation equation allows consistently estimating $g(a;\beta^*)$; the weighted estimation equation for a sample of size $n$, using stabilized weights is:
\begin{equation}\sum_{i=1}^n W_S(A_i,C_i)D(A_i;\beta) [Y_i-g(A_i;\beta)]=0,\label{wgee} \end{equation}
where $D(A_i;\beta)$ is the derivative of $g(A_i;\beta)$ wrt $\beta$.

This result can be obtained in the classical framework, that is without using potential outcome. We wish to estimate the marginal expectation of $Y$ for treatment value $A$ in the experimental situation, $\Ee_{\ex}[Y|A]$.
 This can be done either non-parametrically if $A$ takes a small number of values, or parametrically if $A$ is continuous or can take a large number of values: $\Ee_{\ex}[Y|A=a]=g(a;\beta)$; even if there is one treatment the parameter $\beta$ is multidimensional.
 Acknowledging that $\Ee_{\ex}[Y|A]\ne \Ee_{\ob}[Y|A]$, the problem is that of a change of probability measure in order to relate the model defined in the experimental situation to the observations. The observation probability measure $\PP_{\ob}$ is specified by the joint density of $(Y,A,C)$, $f_{Y,A,C}^{\ob}(y,a,c)$ which we write as
 $$f_{Y,A,C}^{\ob}(y,a,c)=f_{Y|A,C}^{\ob}(y,a|c)f_{A|C}^{\ob}(a|c)f_{C}^{\ob}(c).$$
 Similarly, the joint density under the experimental probability measure is:
 $$f_{Y,A,C}^{\ex}(y,a,c)=f_{Y|A,C}^{\ex}(y,a|c)f_{A|C}^{\ex}(a|c)f_{C}^{\ex}(c).$$
 The physical law is represented by $f_{Y|A,C}$ and does not change under different regime of treatment attribution, so that $f_{Y|A,C}^{\ob}(y,a|c)=f_{Y|A,C}^{\ex}(y,a|c)$. We also choose an experimental situation where $f_{C}^{\ex}(c)=f_{C}^{\ob}(c)$. We can compute the expectation in the experiment from the expectation in the observational probability for any quantity $Q$ as:
 $$\Ee_{\ex}[Q]= \Ee_{\ob}[ZQ],$$
 where $Z$ is a random variable, with the property $\Ee_{\ob}(Z)=1$, taking the value of $\frac{\dd \PP_{\ex}}{\dd \PP_{\ob}}$, the Radon-Nikodym derivative of $\PP_{\ex}$ wrt $\PP_{\ob}$. The Radon-Nikodym derivative exists provided that $\PP_{\ex}$ is absolutely continuous wrt $\PP_{\ob}$; this is related to the so-called ``positivity condition''.
 We have
 $$\frac{\dd \PP_{\ex}}{\dd \PP_{\ob}}=\frac {f_{Y|A,C}^{\ex}(y,a|c)f_{A|C}^{\ex}(a|c)f_{C}^{\ex}(c)}{f_{Y|A,C}^{\ob}(y,a|c)f_{A|C}^{\ob}(a|c)f_{C}^{\ob}(c)}=\frac{f_{A}^{\ex}(a)}{f_{A|C}^{\ob}(a|c)}.$$
If we choose $f_{A}^{\ex}(a)=f_{A}^{\ob}(a)$, we have that $Z=\frac{f_{A}^{\ob}(a)}{f_{A|C}^{\ob}(a|c)}$, which is equal to the stabilized weight. It is not completely obvious that the same variable $Z=W_S$ works for conditional probabilities but it can be verified that
\begin{equation}\Ee_{\ex}[Q|A]= \Ee_{\ob}[ZQ|A].\label{condchange}\end{equation}

Non-parametric inference in the case where $A$ takes a (small) finite number of values by choosing $Q=Y$, a consistent estimator of $\Ee_{\ex}[Y|A=a]$ from an iid sample of $(Y_i,A_i,C_i)$ is
\begin{equation}\frac{1}{\sum_{i=1}^n I(A_i=a)} \sum_{i=1}^n I(A_i=a) Z_i Y_i. \label{marg-simple}\end{equation}
More generally, estimation can be done using the weighted GEE given by Equation (\ref{wgee}). Using Equation (\ref{condchange}) it is shown that this is an unbiased equation in the sense that, for all $i$:
\begin{equation}\Ee_{\ob}\{ W_S(A_i,C_i)D(A_i;\beta) [Y_i-g(A_i;\beta)]|A_i\}=\Ee_{\ex}\{D(A_i;\beta) [Y_i-g(A_i;\beta)]|A_i\}=0.\label{unbiased} \end{equation}
If we apply this equation to the simple case where $A$ can take a small number of values, we retrieve the result of Equation (\ref{marg-simple}). Equation (\ref{unbiased}) allows estimating a parametric model which is necessary when the number of values of $A$ is large.

\citet{murphy2001marginal} used this approach, recognizing that there was no need for counterfactuals.

\subsection{Estimation of the Marginal effect for time-varying exposure with dynamic confounding}\label{sec:dynconfounding}
\citet{robins2000marginal} has tackled the problem of dynamic confounding.
With the aim to avoid the sometimes difficult dynamic modeling, \citet{Robins2000} have proposed the marginal structural models (MSM). However, similarly to the case with simple exposure, marginal structural models can be designed without resorting to potential outcomes. Exactly the same argument as in Section \ref{sec:margeffect}  can be applied and lead to the stabilized weights proposed by \citet{Robins2000}. \citet{Røysland2011} has used this change of measure approach to the more complex continuous case using Girsanov theorem; see \citet{ryalen2018causal} for an application.

Here also there is a natural way to estimate the marginal effect, first estimating a dynamical model, then marginalizing in the law of an experiment, as we have done in Section \ref{sec:throughcond} in the non-dynamical case. Dynamical models have been proposed for instance by \citet{prague2017dynamic}.

\section{The stochastic system approach to causality} \label{sec:SSAC}
\subsection{The stochastic system approach to causality}
In line with the works of \citet{Aalen1987}, \citet{arjas2004causal} and \citet{Didelez2008}, the stochastic system approach to causality has been presented in \citet{Commenges2009}, \citet{commenges2015dynamical} and \citet{commenges2018dealing}. The principle is to define causal influences in a system represented by stochastic processes (rather than random variables), using  the ``local independence'' concept. Intuitively, a process $Y$ is locally independent from a process $V$ if information about $V$ does not change the predicted dynamics of $Y$. If $Y$ is not locally independent of $V$, $V$ directly influences $Y$.
This concept can be defined on a given finite horizon, and importantly, on a random horizon. As in the simple framework of the Section \ref{section:withoutPO}, it is also possible to compute marginal effects, either indirectly from conditional effects, or by a change of measure \citep{Røysland2011}. We do not present again the stochastic system approach in detail here (the reader is referred to the above references), but we give some examples with the aim to show that it is essential to analyze causality in a dynamic framework. One of the problem to which this approach brings a solution is the so-called ``truncation by death''. The approach has also the potential to better understand many epidemiological problems, such as the ``obesity paradox''.

\subsection{Truncation by death}\label{sec:truncationbyD}
 \citet{weuve2015guidelines} have identified the problem  of selection of the sample due to death as one of the major methodological problems for studying risk factors of Alzheimer disease or dementia, or the decrease of cognitive function in the elderly. It is not obvious to treat this problem, which may also arise in other studies of severe diseases, like cancer for instance.
This problem was tackled by \citet{Rubin2006a} using potential outcomes and principal stratification \citep{frangakis2002principal}; the proposed  estimand is the survival average causal effect (SACE). The relevance of principal stratification in some applications has been questioned by  \citet{pearl2011principal} to which  several authors have reacted; in particular, \citet{vanderweele2011principal} noted that the SACE is not identified because it is not known which individuals are in the principal stratum.

\begin{figure}[h!]

\centering
\includegraphics[scale=0.50]{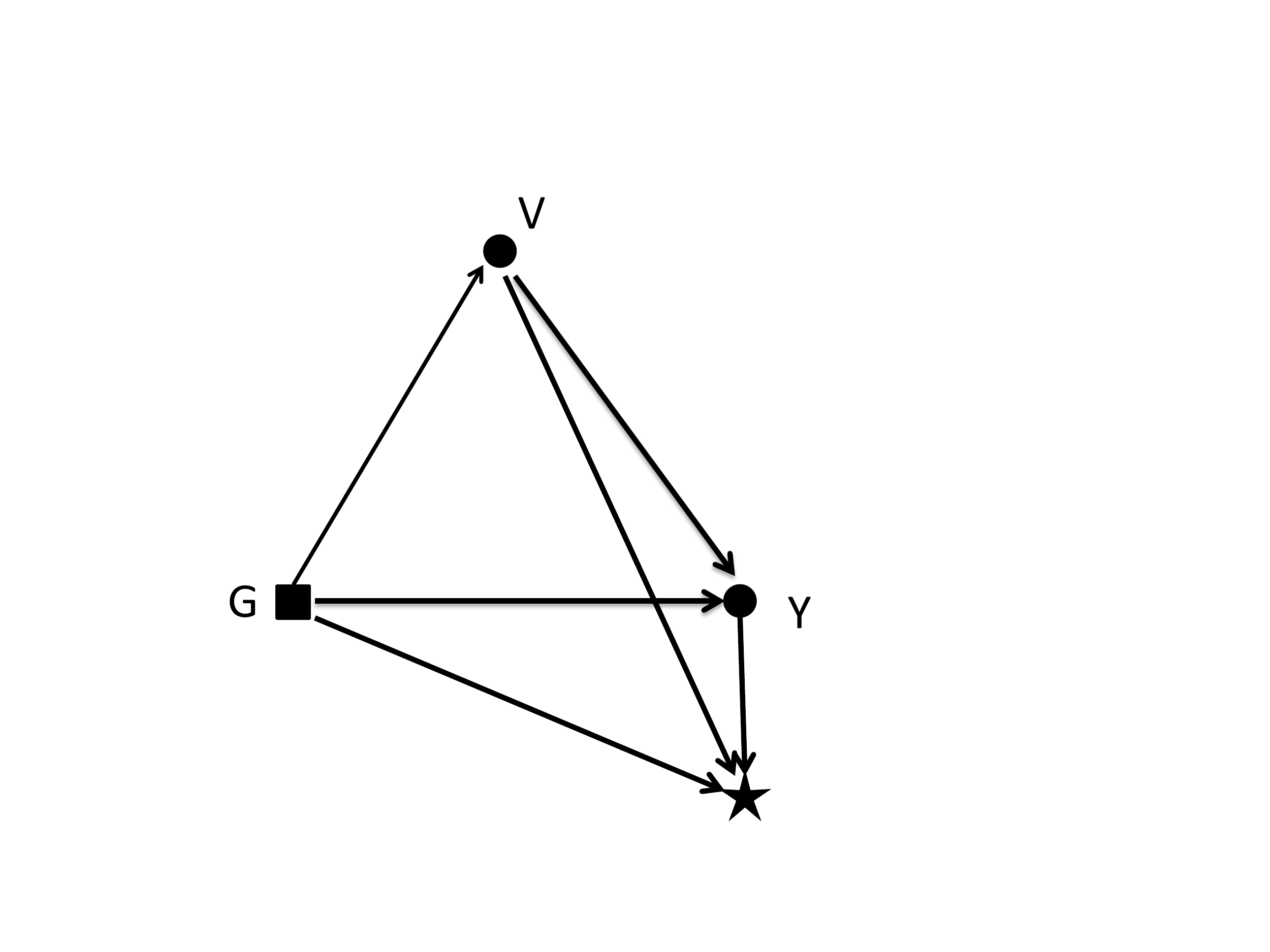}
\caption{Influence graph for a physiological process $Y$, $D$ (death, represented by a star), a
factor $V$ (blood pressure) and attributes $G$ (gender, genetic factors): large filled circles are stochastic processes, and squares are random variables. \label{graphDementia}}
\end{figure}

A solution to this problem in the stochastic system approach to causality has been proposed by \citet{commenges2018dealing}.
The factors of interest are represented by stochastic processes. Using the possibility of studying local influence on a random time, the local influence of a factor $V$ on a marker or disease $Y$ is studied until the time of death. Here, $Y$ could represent dementia or cognitive ability and $V$ blood pressure. A graph of influence between stochastic processes can be built. A typical graph is displayed in Figure \ref{graphDementia}. The graphical convention is that stochastic processes are symbolized by large filled circles, attributes represented by random variables are symbolized by squares and death which is a particular process by a star.

In the principle, this is really straightforward; practical realization of such an analysis relies on developing a careful joint model for $Y$ and death \citep{rizopoulos2010jm,proust-lima_LCMM2015,proust2016joint}.

\subsection{The obesity paradox in Diabetic patients: need for a dynamic approach}
\begin{figure}[h!]
\centering
\includegraphics[scale=0.50]{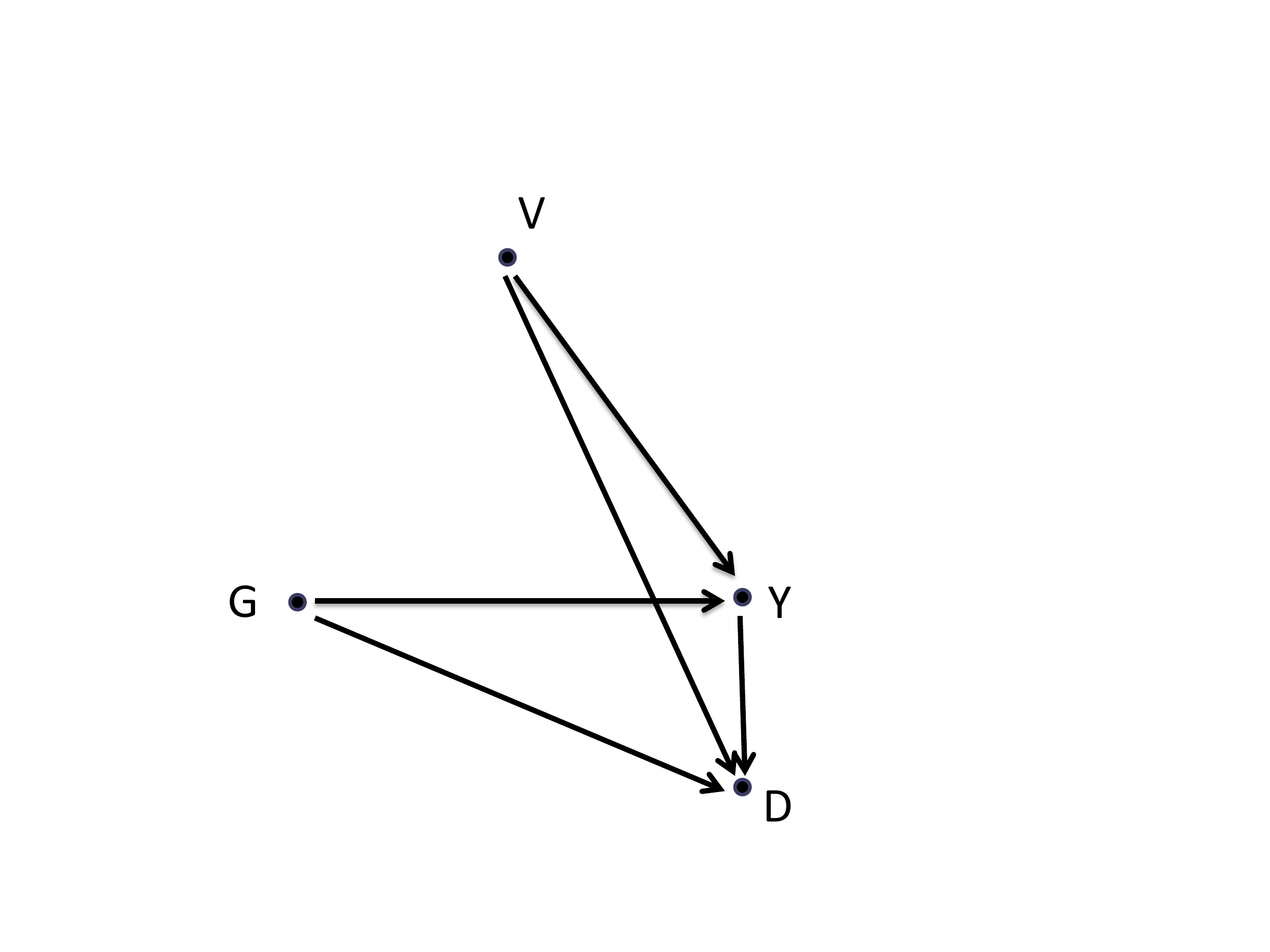}
\caption{Influence graph for  diabetes $Y$, death $D$, obesity $V$ and a possible confounder $G$, all treated as random variables (small filled circles), as in \citet{sperrin2016collider}. \label{graph-Sperrin-diabetes}}
\end{figure}

It is well known that obesity increases the risk of mortality \citep{aune2016bmi} in the general population.
There has been, however, a debate around the so-called ``obesity paradox'' arising from findings that
 in certain populations, obese subjects appear to have a lower death rate; \citet{carnethon2012association} found a hazard ratio of less than 0.5 for overweight/obese subjects. A possible explanation invokes a collider bias, stemming from the fact that many analyzes of mortality among diabetes implicitly condition on diabetes status which may be a collider on the path between obesity and death in presence of a confounder; see Figure \ref{graph-Sperrin-diabetes}. It is, however, controversial that this can explain the paradox \citep{sperrin2016collider,viallon2017re}.

Our point in this problem, is that a modeling through random variables is not well fitted for revealing causal pathways.
Indeed, the two factors studied, obesity and diabetes may vary in time, and death itself is an event which occurs at a certain time, so that all three phenomena are better represented by stochastic processes. A first graphical representation can be the same as in Figure \ref{graphDementia}, where $V$ would represent obesity and $Y$ diabetes: the graph is similar to that of Figure \ref{graph-Sperrin-diabetes} (except for the optional arrow from $G$ to $V$, with the important difference that the nodes are stochastic processes rather than random variables.

\subsubsection{Selection bias}\label{sec:DynSelec}
\begin{figure}[h!]
\centering
\includegraphics[scale=0.50]{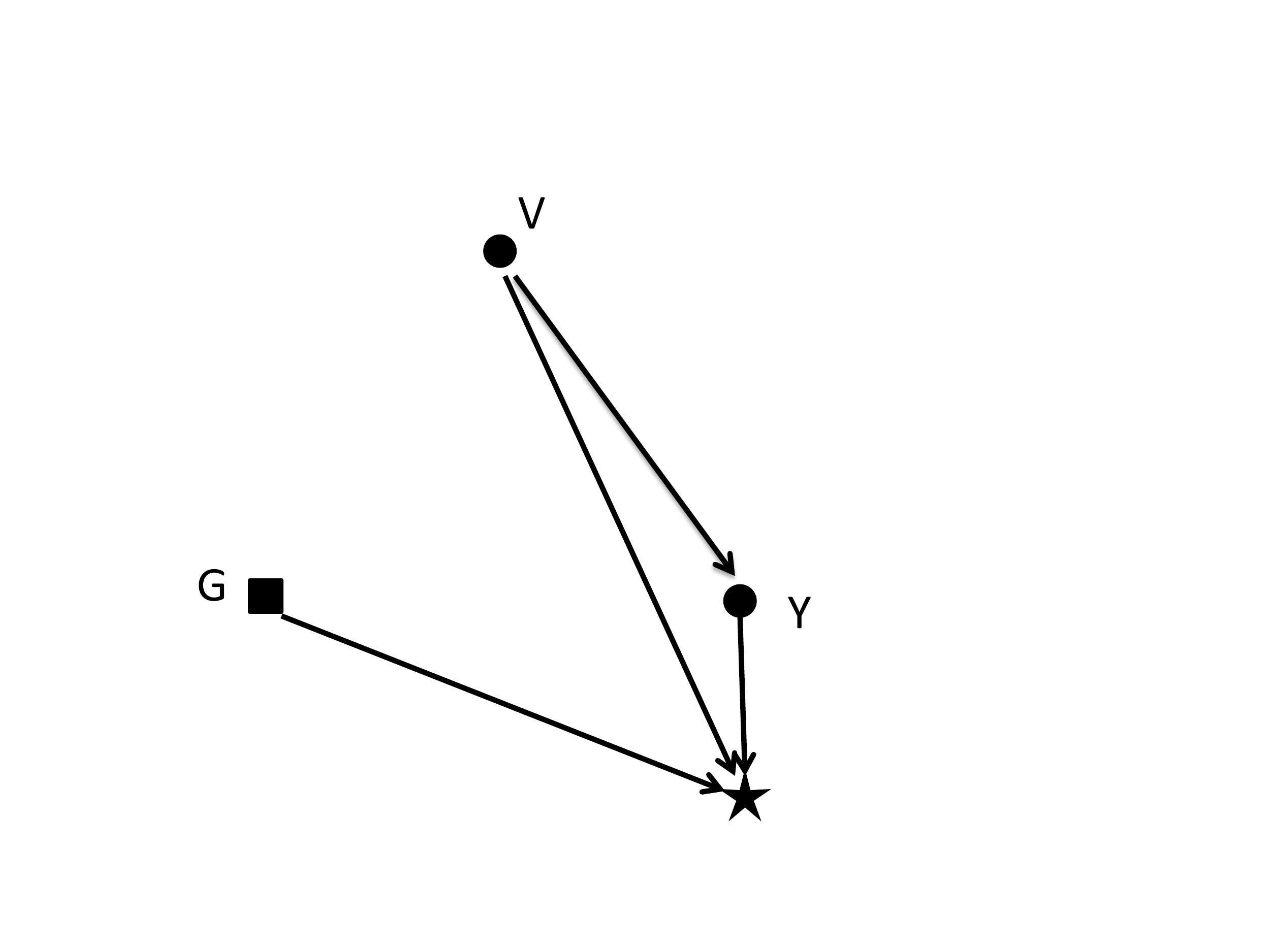}
\caption{Influence graph for  diabetes ($Y$), death (star), obesity ($V$) and a possible confounder ($G$). \label{graph-obesity-simple}}
\end{figure}

In the dynamic version there can be a bias analogous to the collider bias even in the simpler case where $G$ influences only death. This is represented in Figure \ref{graph-obesity-simple}; moreover, in the following discussion we do not use the fact that $V$ influences $Y$ and tha $Y$ influences death.  The potential bias comes from the well known selection of the frailty \citep{aalen2008survival}. Here we assume that $V$ (obesity) is represented by a binary random variable and $G$ takes the role of a frailty. If we assume that conditional on $G$ the hazards are proportional to $V$, marginally the hazard ratio decreases; it may even cross the level of $1$. \citet{aalen2008survival} (Section 6.5.1) studies the case where the frailty has a gamma distribution and the hazard ratio decreases. In such a case, if we start at the time origin, the effect on the survival probability cannot change direction. If, however, we start after or near the time where the hazard ratio crosses the level $1$, we observe an effect in the reverse direction! In the case of obesity, it could happen that obese subjects reaching a certain age are selected to a point that their risk of death is less than non-obese subjects. This may happen if we take a sample of subjects presenting a pathology that happens relatively late, such as type-II diabetes.

\subsubsection{Feedback effect}\label{sec:feedback}
\begin{figure}[h!]
\centering
\includegraphics[scale=0.50]{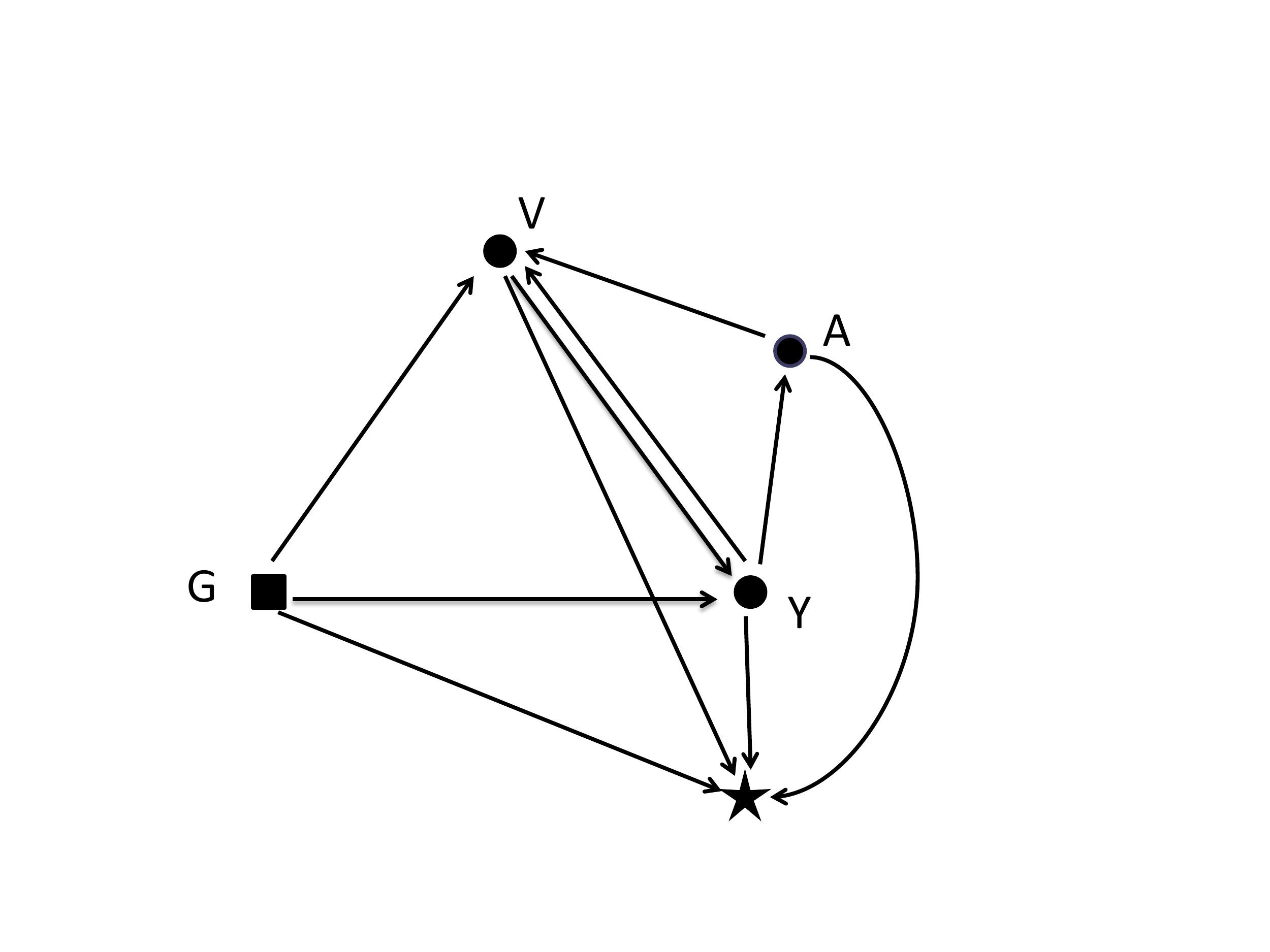}
\caption{Influence graph for diabetes $Y$, death (star), obesity $V$ and a possible confounder $G$; $A$ may represent either a disease or a treatment. \label{graph-obesity-cplx}}
\end{figure}

With the stochastic system approach the influence graph is not necessarily a DAG, that is, is not necessarily acyclic. This allows to represent feedback effects (or inverse causation), which means in our application a direct or indirect effect of diabetes on obesity. The physiological possibility of a direct effect would mean that a severe level of diabetes could induce a loss of weight (possibly by insulino-resistance). Indirect effects are possible through two different mechanisms: (i) diabetes could favor diseases like cancer that could induce weight loss together with increased mortality rate; (ii) a medical intervention for reducing weight could take place in severe diabetes. This latter point is related to the indication bias or ``dynamic confounding'' (see Section \ref{sec:dynconfounding}). So, the influence graph could be as in Figure \ref{graph-obesity-cplx}.
  \citet{knudtson2005associations} have documented both direct and indirect effects of diabetes on obesity showing that ``persons who lost weight had  higher rates of a history of cardiovascular disease and diabete''. Moreover, they reported higher mortality rates in subjects having experienced a weight loss.
The consequence of these biases is that if we take  sample of diabetic patients, the more sever case may have lost weight while they have a higher mortality rate resulting in the ``obesity paradox'', if severity of the diabetes is not taken into account.

For inference we would need longitudinal data, in particular for weight. In such an analysis, the problem of truncation by death would also arise (see Section \ref{sec:truncationbyD}).

\section{Conclusion}\label{sec:conclusion}
 We have shown that we can answer causal questions without potential outcome. Counterfactual questions are not the most interesting but even counterfactual questions can be answered without potential outcome. We have discussed seven issues about the potential outcome approach and shown that causal effects could be defined without it.

 We have put forward the need of a dynamic approach to causality. The issue how to deal with death is important in many epidemiological studies. Only  a dynamic approach can yield a satisfactory solution to this problem. Finally we have examined the example of the obesity paradox. In this problem a dynamic approach allows analyzing different mechanisms that could explain it.
 The two main mechanisms are the frailty selection and reverse causation; the latter can be direct or indirect; the indirect reverse causation could be due either to a disease or a treatment. The reverse causation argument is more convincing if the representation of diabetes (Y) allows to distinguish different levels of severity (either by a quantitative process or a multivariate process). The different shapes of the risk of death for studies of different mean follow-up times found by \citet{aune2016bmi} can be explained by the two mechanisms: less selection and less confounding reverse causation (pre-diagnosis weight-loss) in studies with long follow-up.

 We think that many epidemiological problems would benefit from a dynamic analysis: the obesity paradox occurs in other groups than diabetes;
 the healthy worker effect is also itself a paradox that can be explained dynamically. Dynamical analysis of causality can take two forms: a theoretical analysis as presented here for the obesity paradox; the development of  joint models that would allow making inference from longitudinal data. The two forms should cross-fertilize each-other
\bibliographystyle{chicago}

\bibliography{causality,bibliolivre-trie}

\label{lastpage}

\end{document}